\documentclass[twocolumn,english]{revtex4}
\usepackage[T1]{fontenc}
\usepackage[latin9]{inputenc}
\usepackage{amsmath}
\usepackage{amssymb}
\usepackage{graphicx}

\makeatletter

\providecommand{\tabularnewline}{\\}

\@ifundefined{textcolor}{}
{%
 \definecolor{BLACK}{gray}{0}
 \definecolor{WHITE}{gray}{1}
 \definecolor{RED}{rgb}{1,0,0}
 \definecolor{GREEN}{rgb}{0,1,0}
 \definecolor{BLUE}{rgb}{0,0,1}
 \definecolor{CYAN}{cmyk}{1,0,0,0}
 \definecolor{MAGENTA}{cmyk}{0,1,0,0}
 \definecolor{YELLOW}{cmyk}{0,0,1,0}
 }

\makeatother

\makeatother

\usepackage{babel}
\begin{document}

\title{Quantum repeater with Rydberg blocked atomic ensembles in fiber-coupled
cavities}

\author{E. Brion, F. Carlier, V. M. Akulin }

\address{Laboratoire Aim\'{e} Cotton, CNRS / Univ. Paris-Sud / ENS-Cachan,
B\^{a}t. 505, Campus d'Orsay, 91405 Orsay, France.}

\author{K. Moelmer}

\address{Lundbeck Foundation Theoretical Center for Quantum System Research,
Department of Physics and Astronomy, University of Aarhus, Ny Munkegade,
Bld. 1520, DK-8000, Aarhus C, Denmark. }

\date{{\today } }
\begin{abstract}
We propose and analyze a quantum repeater architecture in which Rydberg
blocked atomic ensembles inside optical cavities are linked by optical
fibers. Entanglement generation, swapping and purification are achieved
through collective laser manipulations of the ensembles and photon
transmission. Successful transmission and storage of entanglement
are heralded by ionization events rather than by photon detection
signal used in previous proposals. We demonstrate how the high charge
detection efficiency allows for a shortened average entanglement generation
time, and we analyze an implementation of our scheme with ensembles
of Cs atoms. 
\end{abstract}


\maketitle

\section{Introduction}

A possible route towards scalable quantum computers and quantum communication
networks combines small quantum processor nodes which communicate
via the exchange of moving information carriers, the so-called \emph{flying
qubits}, which will typically be photons. The direct exchange of a
photon, either through free space or a fiber, between a pair of nodes
does not have unit success probability, and to safely communicate
quantum states, one can instead apply entangled state in teleportation
protocols \cite{B93,B96}, where the entanglement of two remote nodes
can be achieved, e.g. after multiple attempts until a suitable heralding
detection event certifies the establishment of the state. For long
distances, photon loss makes the success probability small, and hence
the average time needed to establish a state for transmission of a
single bit of quantum informsation very long. This problem, however,
can be solved by the quantum repeater setup \cite{BDCZ98} which divides
the transmission path between the nodes into smaller segments with
auxiliary nodes over which losses are strongly diminished. The auxiliary
nodes are first entangled with their nearest neighbors in a heralded
way followed by a succession of local measurements which cause the
graduate projection on quantum states with entanglement distributed
over longer and longer distances (entanglement swapping). The implementation
of this approach is compatible with different physical setups involving
atomic ensembles and linear optical operations. The most influential
proposal is the so-called DLCZ protocol \cite{DLCZ01}, whose feasibility
was experimentally considered in its two-node version in \cite{KBBDCDBK03}.

This manuscript presents a quantum repeater scenario, based on the
Rydberg blockade phenomenon. Rydberg blockade refers to the strong
dipole-dipole interaction between pairs of highly excited atoms, which
after laser excitation of a single atom shifts the resonance condition
for all the other atoms and hence blocks further excitation by the
laser field. Rydberg blockade forbids the resonant excitation of more
than one Rydberg atom in an atomic mesoscopic sample \cite{CP10},
and has been experimentally observed, e.g., in \cite{VVZCCP06,UJHIYWS09}.
In \cite{JCZRCL00} it was proposed to take advantage of Rydberg blockade
in quantum information processing, leading to an intensive current
field of research \cite{SWM10}. Different theoretical proposals have
been recently put forward, which allow to take advantage of the full
spectroscopic richness of Rydberg interactions for quantum information
purposes \cite{BMM07,BPMCS07,BPM07}, and a novel framework for quantum
information encoding and computing has been proposed \cite{LFCD01,BMS07,BPSM08,SWM10},
in which register qubit states are physically implemented by the (symmetric)
occupational states $\left|n_{i}=0,1\right\rangle $ of internal atomic
levels $\left\{ \left|i\right\rangle ,\: i=1,\ldots,K\right\} $ in
an ensemble of $N_{a}$ $(>K$) identical atoms.

The collective laser manipulation of the system combined with the
Rydberg blocking interaction allows one to store and universally process
information in the subspace of symmetric ensemble states containing
at most one atom in each internal level. The primary advantage of
ensembles over single atoms consists in their enhanced coupling to
external control fields, which allows for efficient and rapid processing.
A secondary practical advantage is that even in a multi-qubit register,
one merely needs to address the atoms collectively, contrary to individual-atom
encoding of qubits which requires the precise control of each and
every single particle in the system.

In the quantum repeater we propose here, the nodes are\emph{ $N$}
identical atomic ensembles, placed in cavities which are linked via
optical fibers. The internal structure of the atoms is such that each
ensemble $k=1,\ldots,N$ accomodates three logical subnodes, called
the left $(L_{k})$, the right $(R_{k})$, and the auxiliary subnode
$(A_{k})$ in the following. In the first step of our protocol, we
entangle the logical state of the subnode $(R_{k})$ in the cavity
$k$ with the polarization state of a single photon released in the
cavity. This photon is then transmitted to the neighboring cavity
where it is absorbed by the left subnode $\left(L_{k+1}\right)$ degree
of freedom of the atomic ensemble in that cavity, which thus becomes
entangled with $\left(R_{k}\right)$. A conditional gate applied to
subnodes $(L_{k+1})$ and $(A_{k+1})$, followed by appropriate ionization
detection is used to ensure that no error occurred during the entanglement
generation, in particular that the photon was not lost during the
transfer through the fiber between the cavities. If needed, subnodes
$\left(R_{k},L_{k+1}\right)$ are reset so that the entanglement generation
operation can be repeated until successful. Once all pairs $\left(R_{k},L_{k+1}\right)$
have been correctly entangled, entanglement is swapped by ensemble
operations on every pair $\left(L_{k},R_{k}\right)$. Measurements
using Rydberg blockade and ionization detection on all subnodes $\left\{ L_{k},R_{k},k=2,\ldots,\left(N-1\right)\right\} $
finally heralds the entangled state of the remote pair $\left(R_{1},L_{N}\right)$
which can be transformed into any required Bell state by application
of a unitary operation on $\left(L_{N}\right)$ prescribed by the
results of the measurements. If one of the measurements fails, the
procedure must be repeated.

We note that the use of Rydberg blocked ensembles as quantum repeaters
has been proposed in \cite{ZMHZ10,HHHLS10}. Though related our proposal,
however, never makes use of photon detection to generate entanglement
between neighbouring nodes, we solely rely on ensemble laser manipulations
(including ionizing pulses), ion detection whose efficiency can be
made very close to one $\eta_{d}\sim1$, and photon transmissions
through optical fibers. This allows for a shortened entanglement generation
average time whose expression is derived in the Appendix.

The paper is structured as follows. In Sec. II, we present our quantum
repeater scheme using Rydberg blockaded ensembles in optical cavities
coupled by optical fibers. In Sec. III, we analyze the different steps
of our protocol with emphasis on their robustness against errors,
and we compute the average duration of our scheme. In Sec. IV, we
suggest a physical implementation. In Sec. V, we compare our scheme
with other schemes for quantum repeaters, and we conclude in Sec.
VI.

\section{The model\label{SecII}}

\begin{figure}
\begin{centering}
\includegraphics[width=9cm]{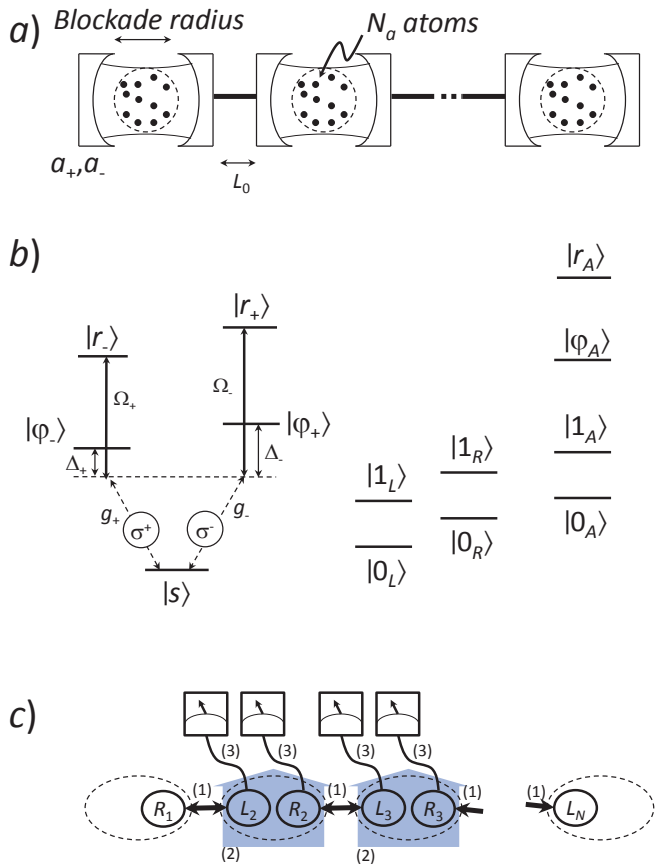} 
\par\end{centering}

\caption{(Color online) a) \emph{The physical set-up for our quantum repeater}.
$N$ atomic ensembles are placed in two-mode cavities linked by optical
fibers. b) \emph{Atomic level structure}. The figure shows the {}``quantum-classical''
two-photon transitions, $\left|s\right\rangle \longleftrightarrow\left|r_{\pm}\right\rangle $
driven via the intermediate states $\left|\varphi_{\pm}\right\rangle $
with coupling strength $g_{\pm}$ to the quantized cavity field modes
$\pm$ and with Rabi frequencies $\Omega_{\pm}$ detunings $\Delta_{\pm}$
to the classical laser control fields. c) \emph{The three steps of
our scheme}. (1) entanglement of pairs of neighboring subnodes $\left(R_{k},L_{k+1}\right)$,
entanglement swapping through two-bit gates (2) on each pair $\left(L_{k},R_{k}\right)$
and measurement (3) of all subnodes but $\left(R_{1},L_{N}\right)$.}

\label{Fig1} 
\end{figure}

Our quantum repeater setup consists of $N$ atomic ensembles placed
in cavities which are linked by optical fibers (see Fig. \ref{Fig1}a).
Neighboring cavities are separated by the distance $L_{0}$.

The atomic level structure is represented in Fig. \ref{Fig1}b. All
the atoms are initially prepared in the {}``reservoir'' state $\left|s\right\rangle $,
and the atoms have, in addition, six metastable states denoted by
$\left|0_{L}\right\rangle ,\left|1_{L}\right\rangle ,\left|0_{R}\right\rangle ,\left|1_{R}\right\rangle ,\left|0_{A}\right\rangle ,\left|1_{A}\right\rangle $,
three excited states $\left|\varphi_{+}\right\rangle ,\left|\varphi_{-}\right\rangle ,\left|\varphi_{A}\right\rangle $
and three high-lying Rydberg states $\left|r_{+}\right\rangle ,\left|r_{-}\right\rangle ,\left|r_{A}\right\rangle $.
We assume that the transitions given in Table \ref{Transitions} can
be independently and selectively addressed by appropriately tuned
laser beams. In particular, it implies that we can couple pairs of
states, $\left|s\right\rangle \leftrightarrow\left|r_{\pm,A}\right\rangle $,
$\left|0_{L,R}\right\rangle \leftrightarrow\left|r_{-}\right\rangle $,
$\left|1_{L,R}\right\rangle \leftrightarrow\left|r_{+}\right\rangle $
and $\left|0_{A}\right\rangle ,\left|1_{A}\right\rangle \leftrightarrow\left|r_{A}\right\rangle $
via the intermediate states $\left|\varphi_{+}\right\rangle $, $\left|\varphi_{-}\right\rangle $
or $\left|\varphi_{A}\right\rangle $.

We further suppose that the atomic samples are small enough to operate
in the full Rydberg blockade regime, $i.e.$ their size should not
exceed a few $\mu$m. As a consequence, when driving the transition
$\left|s\right\rangle \leftrightarrow\left|r_{\pm,A}\right\rangle $
on a sample with $N_{a}$ atoms initially in the state $\left|s\ldots s\right\rangle $,
multiply excited states are out of resonance due to the strong dipole-dipole
interaction among Rydberg excited atoms, and the transfer of more
than a single atom to the Rydberg state $\left|r_{\pm,A}\right\rangle $
is blocked. The fields are applied symmetrically to all atoms in each
sample, and they hence excite the symmetric collective state with
a single Rydberg excitation, $\left(\left|r_{k}s\ldots s\right\rangle +\left|sr_{k}s\ldots s\right\rangle +\ldots+\left|s\ldots sr_{k}\right\rangle \right)/\sqrt{N_{a}}$.
The associated coupling strength is easily seen to be magnified by
the factor $\sqrt{N_{a}}$ with respect to the coupling strength of
the single atom transition $\left|s\right\rangle \leftrightarrow\left|r_{k}\right\rangle $.
Applying a $\pi$ pulse on the collective ensemble transition, followed
by the single-particle transition $\left|r_{k}\right\rangle \rightarrow\left|j\right\rangle $
$\left(j=0_{L},1_{L},0_{R},1_{R},0_{A},1_{A}\right)$, prepares the
sample in a stable symmetric collective state $\left|N_{0_{L}},N_{1_{L}},N_{0_{R}},N_{1_{R}},N_{0_{A}},N_{1_{A}}\right\rangle $
where the $N_{j}$'s denote the populations of the different internal
levels, restricted to values $0$ and $1$. Unitary operations can
be applied in the eight-dimensional subspace of collective states,
$\left\{ \left|N_{0_{L}},\left(1-N_{0_{L}}\right),N_{0_{R}},\left(1-N_{0_{R}}\right),N_{0_{A}},\left(1-N_{0_{A}}\right)\right\rangle \right\} $,
by simply driving the corresponding single-atom transitions $\left|0_{L,A,R}\right\rangle \leftrightarrow\left|1_{L,A,R}\right\rangle $
and/or $\left|0_{L,A,R}\right\rangle ,\left|1_{L,A,R}\right\rangle \leftrightarrow\left|r_{+,-,R}\right\rangle $,
as described in \cite{BMS07,BPM07}. The collective state pairs ($\left|0_{L}\right\rangle ,\left|1_{L}\right\rangle $),
($\left|0_{R}\right\rangle ,\left|1_{R}\right\rangle $), and ($\left|0_{A}\right\rangle ,\left|1_{A}\right\rangle $)
at each repeater node can thus be associated with three qubits, referred
to as the left $\left(L\right)$, right $\left(R\right)$ and auxiliary
$\left(A\right)$ subnodes, in the following.

We shall use a simplified notation for ensemble states, denoting the
collective internal state population as $\bar{0}$ and $\bar{1}$,
such that, e.g., $\left|\bar{0}_{L};\bar{1}_{R};\bar{1}_{A}\right\rangle $
denotes the state $\left|N_{0_{L}}=N_{1_{R}}=N_{1_{A}}=1;\, N_{1_{L}}=N_{0_{R}}=N_{0_{A}}=0\right\rangle $,
where both the right and the auxiliary subnodes occupy the logic state
1, while the left subnode is in state $0$.

\begin{table}
\begin{centering}
\begin{tabular}{|c|}
\hline 
$\left|0,1_{A}\right\rangle \leftrightarrow\left|\varphi_{A}\right\rangle $\tabularnewline
\hline 
$\left|0_{L,R}\right\rangle \leftrightarrow\left|\varphi_{-}\right\rangle $\tabularnewline
\hline 
$\left|1_{L,R}\right\rangle \leftrightarrow\left|\varphi_{+}\right\rangle $\tabularnewline
\hline 
$\left|s\right\rangle \overset{\sigma_{\pm}}{\longleftrightarrow}\left|\varphi_{\pm}\right\rangle $\tabularnewline
\hline 
$\left|\varphi_{A}\right\rangle \leftrightarrow\left|r_{A}\right\rangle $\tabularnewline
\hline 
$\left|\varphi_{\pm}\right\rangle \leftrightarrow\left|r_{\pm}\right\rangle $\tabularnewline
\hline 
$\left|0_{L,R,A}\right\rangle \leftrightarrow\left|1_{L,R,A}\right\rangle $\tabularnewline
\hline 
\end{tabular}
\par\end{centering}

\caption{The transitions required by our protocol.}

\label{Transitions} 
\end{table}

It is a further requirement of our protocol that the transitions $\left|s\right\rangle \longleftrightarrow\left|\varphi_{\pm}\right\rangle $
couple non-resonantly to two different modes, $\pm$, with equal frequency
and with detunings $\Delta_{\pm}$ with respect to the atomic transitions,
but different polarization in the cavity (this will legitimate the
assumptions we make below on the fiber transmission/losses). Driving
the upper transition $\left|r_{\pm}\right\rangle \longleftrightarrow\left|\varphi_{\pm}\right\rangle $
with a laser field with detuning $\Delta_{\pm}$ and Rabi frequency
$\Omega_{\pm}$, induces a second-order quantum-classical process,
described by the single-atom effective Hamiltonian $\frac{\hbar g_{\pm}\Omega_{\pm}}{\Delta_{\pm}}a_{\pm}^{\dagger}\left|s\right\rangle \left\langle r_{\pm}\right|+h.c.$
where $a_{\pm}$ denotes the annihilation operator of the cavity mode
$\pm$ and $g_{\pm}$ its coupling strength. This Hamiltonian derives
from the adiabatic elimination of the intermediate state $\left|\varphi_{\pm}\right\rangle $
\cite{BPM07-1}, and is valid for $\left|\Delta_{\pm}\right|\gg\left|\Omega_{\pm}\right|,\left|g_{\pm}\right|$.
Performing an ensemble $\pi$ pulse on the quantum-classical two-photon
transition $\left|s\right\rangle \leftrightarrow\left|\varphi_{\pm}\right\rangle \leftrightarrow\left|r_{\pm}\right\rangle $
hence converts a collective Rydberg excitation $\left|r_{\pm}\right\rangle $
into a cavity photon $\pm$ and the other way around, with the coupling
strength \textbf{$\hbar\sqrt{N_{a}}g_{\pm}\Omega_{\pm}/\Delta_{\pm}$}.
We emphasize that, though only a \emph{single} cavity photon is emitted/absorbed
during the process, the coupling can be made strong, thanks to the
atomic ensemble magnification factor \textbf{$\sqrt{N_{a}}$} \cite{GBEM10}.
To avoid spurious interference effects we suggest to apply different
detuning parameters for the two transitions proceeding via the same
intermediate state $\left|\varphi_{\pm}\right\rangle $.

Optical fibers \cite{P97} couple the cavities to each other, and
we simply assume that the fiber between two neighboring cavities $\left(k,k+1\right)$
achieves the coupling 
\begin{equation}
V_{k,k+1}=\sum_{j=0,1}\hbar\alpha_{k,j}\left[a_{k,j}a_{k+1,j}^{\dagger}+a_{k,j}^{\dagger}a_{k+1,j}\right]\label{FiberCoupling}
\end{equation}
 with a coupling strength $\alpha_{k,j}$ between the modes $j=+,-$
in the $k^{th}$ cavity with annihilation operator $a_{k,j}$, and
the same modes in the $\left(k+1\right)^{th}$ cavity. Since the modes
$\left(+,-\right)$ differ by their polarizations but have the same
frequency, we make the reasonable assumption that the $\alpha_{k,j}$'s
have the same value $\alpha$ for all $\left(k,j\right)$. We moreover
suppose that the coupling between two neighboring cavities can be
switched on and off, for instance by a controlled Pockels cell : this
will allow us to isolate and separately deal with pairs of coupled
nodes.

Due to fiber loss, the transmission of a photon from one cavity to
its neighbor is not perfect. We shall assume that the transmission
efficiency is the same for all fiber connections and can be written
under the form $\eta_{t}=\exp\left(-L_{0}/L_{att}\right)$ where $L_{att}$
is the attenuation length, typically of the order of $22$km (corresponding
to losses of $0.2$dB/km). The probability for loosing one photon
$\left(+,-\right)$ during the transmission along a fiber mode of
length $L_{0}$ is thus given by $\left(1-\eta_{t}\right)$.

\section{The scheme \label{sec:The-scheme}}

In this section, we describe how to entangle two remote nodes using
the model we presented in the previous section. First, we briefly
sketch the different steps of our scheme in the ideal case without
losses. Then, we show how to make our method immune to photon loss
and spontaneous emission from the Rydberg level. Finally, we analyze
the effects of other possible errors on the performance of our protocol.

\subsection{The different steps of the scheme}

Initially, all the cavities are empty and all ensemble atoms are in
the reservoir state $\left|s\ldots s\right\rangle $. The first step
consists in entangling pairs of neighboring subnodes $\left(R_{k},L_{k+1}\right)$
(see Fig. \ref{Fig1}c). To this end, one applies the sequence of
operations given in Table \ref{EntanglementGeneration}. In this table,
dashed-line arrows are used for non-resonant couplings to the intermediate
states. Moreover, the {}``ensemble'' nature of pulses is emphasized
by a $\sqrt{N_{a}}$ factor above the concerned arrows. Finally, we
indicate when a cavity mode is involved in a transition $\left|s\right\rangle \leftrightarrow\left|\varphi_{\pm}\right\rangle $
by writing {}``single photon $\pm$'' above the corresponding arrow.

\begin{widetext}

\begin{table}
\begin{centering}
\begin{tabular}{|l|}
\hline 
\emph{i}) simultaneous $\pi$ pulses (same Rabi frequency) on ensemble
$\left(k\right)$\tabularnewline
$\left\{ \begin{array}{c}
\left|s\right\rangle \overset{\sqrt{N_{a}}}{\dashrightarrow}\left|\varphi_{-}\right\rangle \dashrightarrow\left|r_{-}\right\rangle \\
\left|s\right\rangle \overset{\sqrt{N_{a}}}{\dashrightarrow}\left|\varphi_{+}\right\rangle \dashrightarrow\left|r_{+}\right\rangle 
\end{array}\right.$\tabularnewline
\hline 
\emph{ii}) $\pi$ pulse on ensemble $\left(k\right)$ $\left|r_{-}\right\rangle \dashrightarrow\left|\varphi_{-}\right\rangle \overset{\textrm{single photon -, }\sqrt{N_{a}}}{\dashrightarrow}\left|s\right\rangle $\tabularnewline
\hline 
\emph{iii}) $\pi$ pulse on ensemble $\left(k\right)$ $\left|s\right\rangle \overset{\sqrt{N_{a}}}{\dashrightarrow}\left|\varphi_{-}\right\rangle \dashrightarrow\left|r_{-}\right\rangle $\tabularnewline
\hline 
\emph{iv}) $\pi$ pulse on ensemble $\left(k\right)$ $\left|r_{+}\right\rangle \dashrightarrow\left|\varphi_{+}\right\rangle \overset{\textrm{single photon +, }\sqrt{N_{a}}}{\dashrightarrow}\left|s\right\rangle $\tabularnewline
\hline 
\emph{v}) $\pi$ pulse on ensemble $\left(k\right)$ $\left|s\right\rangle \overset{\sqrt{N_{a}}}{\dashrightarrow}\left|\varphi_{+}\right\rangle \dashrightarrow\left|r_{+}\right\rangle $\tabularnewline
\hline 
\emph{vi}) $\pi$ pulse on ensemble $\left(k\right)$ $\left|r_{-}\right\rangle \dashrightarrow\left|\varphi_{-}\right\rangle \dashrightarrow\left|0_{R}\right\rangle $\tabularnewline
\hline 
\emph{vii}) $\pi$ pulse on ensemble $\left(k\right)$ $\left|r_{+}\right\rangle \dashrightarrow\left|\varphi_{+}\right\rangle \dashrightarrow\left|1_{R}\right\rangle $\tabularnewline
\hline 
\emph{viii}) \textbf{Transfer} \textbf{of the photon} through the
fiber \tabularnewline
from cavity $\left(k\right)$ to cavity $\left(k+1\right)$ \tabularnewline
\hline 
\emph{ix}) $\pi$ pulse on ensemble $\left(k+1\right)$ $\left|s\right\rangle \overset{\textrm{single photon -, }\sqrt{N_{a}}}{\dashrightarrow}\left|\varphi_{-}\right\rangle \dashrightarrow\left|r_{-}\right\rangle $\tabularnewline
\hline 
\emph{x}) $\pi$ pulse on ensemble $\left(k+1\right)$ $\left|r_{-}\right\rangle \dashrightarrow\left|\varphi_{-}\right\rangle \dashrightarrow\left|0_{L}\right\rangle $\tabularnewline
\hline 
\emph{xi}) $\pi$ pulse on ensemble $\left(k+1\right)$ $\left|s\right\rangle \overset{\textrm{single photon +, }\sqrt{N_{a}}}{\dashrightarrow}\left|\varphi_{+}\right\rangle \dashrightarrow\left|r_{+}\right\rangle $\tabularnewline
\hline 
\emph{xii}) $\pi$ pulse on ensemble $\left(k+1\right)$ $\left|r_{+}\right\rangle \dashrightarrow\left|\varphi_{+}\right\rangle \dashrightarrow\left|1_{L}\right\rangle $\tabularnewline
\hline 
\end{tabular}
\par\end{centering}

\caption{The different steps of the entanglement generation procedure.}

\label{EntanglementGeneration} 
\end{table}

\end{widetext}

The first seven pulses $\left(i-vii\right)$ in Table \ref{EntanglementGeneration}
prepare the subnode $\left(R_{k}\right)$ and the cavity $\left(k\right)$
in the entangled state $\left(\left|\bar{0}_{R}\right\rangle \otimes\left|-\right\rangle +\left|\bar{1}_{R}\right\rangle \otimes\left|+\right\rangle \right)/\sqrt{2}$
where $\left|\pm\right\rangle $ denotes the number state $\left|n_{\pm}=1\right\rangle $
of the cavity mode $\pm$. The photon thus released is then transferred
to the cavity $\left(k+1\right)$ through the fiber (step $viii$).
The last four pulses $\left(ix-xii\right)$ of the sequence translate
the photonic excitation into an atomic excitation of the ensemble
$\left(k+1\right)$ : a {}``$-$'' photon is translated into a $\left|0_{L}\right\rangle $
excitation, a {}``$+$'' photon into a $\left|1_{L}\right\rangle $
excitation. Finally, the two subnodes $\left(R_{k},L_{k+1}\right)$
are left in the state $\left(\left|\bar{0}_{R}\right\rangle _{k}\otimes\left|\bar{0}_{L}\right\rangle _{k+1}+\left|\bar{1}_{R}\right\rangle _{k}\otimes\left|\bar{1}_{L}\right\rangle _{k+1}\right)/\sqrt{2}$.
The sequence of states along which the system evolves during the series
of operations described in Table \ref{EntanglementGeneration} can
be found in Table \ref{EntanglementGenerationStates}.

\begin{widetext}

\begin{table}
\begin{centering}
\begin{tabular}{|l|}
\hline 
$\left|s\ldots s\right\rangle _{k}\otimes\left|\textrm{vac}\right\rangle _{k}\otimes\left|s\ldots s\right\rangle _{k+1}\otimes\left|\textrm{vac}\right\rangle _{k+1}$\tabularnewline
\hline 
$\quad\overset{i)}{\rightarrow}\left(\left|N_{r_{-}}=1\right\rangle _{k}+\left|N_{r_{+}}=1\right\rangle _{k}\right)/\sqrt{2}\otimes\left|\textrm{vac}\right\rangle _{k+1}\otimes\left|s\ldots s\right\rangle _{k+1}\otimes\left|\textrm{vac}\right\rangle _{k+1}$\tabularnewline
\hline 
$\quad\overset{ii)}{\rightarrow}\left(\left|s\ldots s\right\rangle _{k}\otimes\left|-\right\rangle _{k}+\left|N_{r_{+}}=1\right\rangle _{k}\otimes\left|\textrm{vac}\right\rangle _{k}\right)/\sqrt{2}\otimes\left|s\ldots s\right\rangle _{k+1}\otimes\left|\textrm{vac}\right\rangle _{k+1}$\tabularnewline
\hline 
$\quad\overset{iii)}{\rightarrow}\left(\left|N_{r_{-}}=1\right\rangle _{k}\otimes\left|-\right\rangle _{k}+\left|N_{r_{+}}=1\right\rangle _{k}\otimes\left|\textrm{vac}\right\rangle _{k}\right)/\sqrt{2}\otimes\left|s\ldots s\right\rangle _{k+1}\otimes\left|\textrm{vac}\right\rangle _{k+1}$\tabularnewline
\hline 
$\quad\overset{iv)}{\rightarrow}\left(\left|N_{r_{-}}=1\right\rangle _{k}\otimes\left|-\right\rangle _{k}+\left|s\ldots s\right\rangle _{k}\otimes\left|+\right\rangle _{k}\right)/\sqrt{2}\otimes\left|s\ldots s\right\rangle _{k+1}\otimes\left|\textrm{vac}\right\rangle _{k+1}$\tabularnewline
\hline 
$\quad\overset{v)}{\rightarrow}\left(\left|N_{r_{-}}=1\right\rangle _{k}\otimes\left|-\right\rangle _{k}+\left|N_{r_{+}}=1\right\rangle _{k}\otimes\left|+\right\rangle _{k}\right)/\sqrt{2}\otimes\left|s\ldots s\right\rangle _{k+1}\otimes\left|\textrm{vac}\right\rangle _{k+1}$\tabularnewline
\hline 
$\quad\overset{vi)}{\rightarrow}\left(\left|\bar{0}_{R}\right\rangle _{k}\otimes\left|-\right\rangle _{k}+\left|N_{r_{+}}=1\right\rangle _{k}\otimes\left|+\right\rangle _{k}\right)/\sqrt{2}\otimes\left|s\ldots s\right\rangle _{k+1}\otimes\left|\textrm{vac}\right\rangle _{k+1}$\tabularnewline
\hline 
$\quad\overset{vii)}{\rightarrow}\left(\left|\bar{0}_{R}\right\rangle _{k}\otimes\left|-\right\rangle _{k}+\left|\bar{1}_{R}\right\rangle _{k}\otimes\left|+\right\rangle _{k}\right)/\sqrt{2}\otimes\left|s\ldots s\right\rangle _{k+1}\otimes\left|\textrm{vac}\right\rangle _{k+1}$\tabularnewline
\hline 
$\quad\overset{viii)}{\rightarrow}\left[\begin{array}{c}
\left|\bar{0}_{R}\right\rangle _{k}\otimes\left|\textrm{vac}\right\rangle _{k}\otimes\left|s\ldots s\right\rangle _{k+1}\otimes\left|-\right\rangle _{k+1}\\
+\left|\bar{1}_{R}\right\rangle _{k}\otimes\left|\textrm{vac}\right\rangle _{k}\otimes\left|s\ldots s\right\rangle _{k+1}\otimes\left|+\right\rangle _{k+1}
\end{array}\right]/\sqrt{2}$\tabularnewline
\hline 
$\quad\overset{ix)}{\rightarrow}\left[\begin{array}{c}
\left|\bar{0}_{R}\right\rangle _{k}\otimes\left|\textrm{vac}\right\rangle _{k}\otimes\left|N_{r_{-}}=1\right\rangle _{k+1}\otimes\left|\textrm{vac}\right\rangle _{k+1}\\
+\left|\bar{1}_{R}\right\rangle _{k}\otimes\left|\textrm{vac}\right\rangle _{k}\otimes\left|s\ldots s\right\rangle _{k+1}\otimes\left|+\right\rangle _{k+1}
\end{array}\right]/\sqrt{2}$\tabularnewline
\hline 
$\quad\overset{x)}{\rightarrow}\left[\begin{array}{c}
\left|\bar{0}_{R}\right\rangle _{k}\otimes\left|\textrm{vac}\right\rangle _{k}\otimes\left|\bar{0}_{L}\right\rangle _{k+1}\otimes\left|\textrm{vac}\right\rangle _{k+1}\\
+\left|\bar{1}_{R}\right\rangle _{k}\otimes\left|\textrm{vac}\right\rangle _{k}\otimes\left|s\ldots s\right\rangle _{k+1}\otimes\left|+\right\rangle _{k+1}
\end{array}\right]/\sqrt{2}$\tabularnewline
\hline 
$\quad\overset{xi)}{\rightarrow}\left[\begin{array}{c}
\left|\bar{0}_{R}\right\rangle _{k}\otimes\left|\textrm{vac}\right\rangle _{k}\otimes\left|\bar{0}_{L}\right\rangle _{k+1}\otimes\left|\textrm{vac}\right\rangle _{k+1}\\
+\left|\bar{1}_{R}\right\rangle _{k}\otimes\left|\textrm{vac}\right\rangle _{k}\otimes\left|N_{r_{+}}=1\right\rangle _{k+1}\otimes\left|\textrm{vac}\right\rangle _{k+1}
\end{array}\right]/\sqrt{2}$\tabularnewline
\hline 
$\quad\overset{xii)}{\rightarrow}\left[\begin{array}{c}
\left|\bar{0}_{R}\right\rangle _{k}\otimes\left|\textrm{vac}\right\rangle _{k}\otimes\left|\bar{0}_{L}\right\rangle _{k+1}\otimes\left|\textrm{vac}\right\rangle _{k+1}\\
+\left|\bar{1}_{R}\right\rangle _{k}\otimes\left|\textrm{vac}\right\rangle _{k}\otimes\left|\bar{1}_{L}\right\rangle _{k+1}\otimes\left|\textrm{vac}\right\rangle _{k+1}
\end{array}\right]/\sqrt{2}$\tabularnewline
\hline 
\end{tabular}
\par\end{centering}

\caption{Evolution of the state vector during the entanglement generation procedure.}

\label{EntanglementGenerationStates} 
\end{table}

\end{widetext}

The entanglement generation procedure described above cannot be applied
simultaneously on all pairs $\left(R_{k},L_{k+1}\right)$ : indeed,
the pair of nodes to be entangled must be isolated from the others
for the photon exchange. One can, however, deal with all the pairs
$\left(R_{2k-1},L_{2k}\right)_{1\leq k\leq N/2}$ in parallel. Once
entanglement has been successfully established among these pairs,
one can then treat the remaining pairs $\left(R_{2k},L_{2k+1}\right)_{1\leq k\leq N/2-1}$.
Omitting the auxiliary subnodes and the cavity modes which are all
empty, one can write the final state of the system under the form
$\prod_{k=1}^{N-1}\left(\left|\bar{0}_{R}\right\rangle _{k}\otimes\left|\bar{0}_{L}\right\rangle _{k+1}+\left|\bar{1}_{R}\right\rangle _{k}\otimes\left|\bar{1}_{L}\right\rangle _{k+1}\right)/\sqrt{2}$.\textbf{ }

To complete the scheme, we now need to swap entanglement, \emph{i.e.}
to entangle the left and right subnodes $\left(R_{k}\right)$ and
$\left(L_{k}\right)$ in every ensemble and decouple the first and
last nodes from all the others. This constitutes the second step of
the method (see Fig. \ref{Fig1} c). To this end, one first simultaneously
applies to each pair of subnodes $\left(R_{k},L_{k}\right)_{k=2,\ldots,N-1}$
the unitary transformation $\left(U_{L}\otimes U_{R}\right)\times P_{LR}\times\left(\mathbb{I}_{L}\otimes V_{R}\right)$,
where $U=\exp\left(-\mathrm{i}\frac{\pi}{2}\sigma_{z}\right)\times\exp\left(-\mathrm{i}\frac{\pi}{4}\sigma_{y}\right)$
and $V=\exp\left(-\mathrm{i}\frac{\pi}{2}\sigma_{x}\right)\times\exp\left(-\mathrm{i}\frac{\pi}{2}\sigma_{z}\right)\times\exp\left(-\mathrm{i}\frac{\pi}{4}\sigma_{y}\right)$
are simply achieved through applying the appropriate laser-induced
pulses $\left|0_{L,R}\right\rangle \leftrightarrow\left|1_{L,R}\right\rangle $
and $P_{LR}=\left(\begin{array}{cccc}
-1 & 0 & 0 & 0\\
0 & -1 & 0 & 0\\
0 & 0 & 1 & 0\\
0 & 0 & 0 & -1
\end{array}\right)$ is implemented through the following sequence of pulses : 
\begin{eqnarray*}
\pi\textrm{ pulse :}\left|0_{L}\right\rangle  & \rightarrow & \left|r_{-}\right\rangle \\
2\pi\textrm{ pulse :}\left|1_{R}\right\rangle  & \longleftrightarrow & \left|r_{+}\right\rangle \\
\pi\textrm{ pulse :}\left|r_{-}\right\rangle  & \rightarrow & \left|0_{L}\right\rangle .
\end{eqnarray*}
 (Note that all these processes are driven by classical laser beams,
via the intermediate states $\left|\varphi_{\pm}\right\rangle $.)
Finally one measures all subnodes $\left(R_{k},L_{k}\right)_{k=2,\ldots,N-1}$
through state-selective ionization. This step can be achieved in parallel
on the different subnodes. The average time needed for the entanglement
swapping operation is hence the time needed for performing the gate
and the measurement on a single ensemble.

At the end of the whole procedure, the subnodes $\left(R_{1}\right)$
and $\left(L_{N}\right)$, are decoupled from all the others and reduced
in one of the four entangled states $\left\{ \left(\left|00\right\rangle \pm\left|11\right\rangle \right)/\sqrt{2},\left(\left|01\right\rangle \pm\left|10\right\rangle \right)/\sqrt{2}\right\} $.
The unitary $W$ one has to apply on the qubit stored in $\left(L_{N}\right)$
-- through driving the appropriate pulse $\left|0_{L}\right\rangle \leftrightarrow\left|1_{L}\right\rangle $,
to get the desired state $\left(\left|00\right\rangle +\left|11\right\rangle \right)/\sqrt{2}$
is determined from the outcomes of the measurements : it is indeed
obtained as the product of $\left(N-2\right)$ transformations $W=\prod_{k=2}^{N-1}W_{k}$,
where $W_{k}$ depends on the values $\left(i_{L_{k}},i_{R_{k}}\right)_{k=2,\ldots,\left(N-1\right)}$
found for the qubits stored in the left and right subnodes of the
ensemble $\left(k\right)$: 
\begin{eqnarray*}
\left(i_{L_{k}},i_{R_{k}}\right) & = & \left(0,0\right)\Rightarrow W_{k}=I\\
\left(i_{L_{k}},i_{R_{k}}\right) & = & \left(0,1\right)\Rightarrow W_{k}=\sigma_{x}\\
\left(i_{L_{k}},i_{R_{k}}\right) & = & \left(1,0\right)\Rightarrow W_{k}=\sigma_{z}\\
\left(i_{L_{k}},i_{R_{k}}\right) & = & \left(1,1\right)\Rightarrow W_{k}=\sigma_{z}\sigma_{x}
\end{eqnarray*}
 where $\sigma_{x}\equiv\left(\begin{array}{cc}
0 & 1\\
1 & 0
\end{array}\right)$ and $\sigma_{z}\equiv\left(\begin{array}{cc}
1 & 0\\
0 & -1
\end{array}\right)$ are the usual Pauli matrices.

\subsection{Error detection and prevention}

So far, we did not take into account errors and losses. Fiber loss
and spontaneous emission from the Rydberg level may corrupt the state
in quite the same way since they both represent the loss of an excitation.
We now investigate the influence of such errors on the different steps
of our scheme.

If a photon loss occurs during the transfer through the fiber or if
a Rydberg excited atom spontaneously decays to the reservoir state
during one of the steps in Table \ref{EntanglementGeneration}, an
excitation is missing either in $\left(R_{k}\right)$ or $\left(L_{k+1}\right)$
at the end of the entanglement generation procedure. To diagnose whether
this is the case, we merely need to test the occupancy of both subspaces
$\left\{ \left|0_{R}\right\rangle _{k},\left|1_{R}\right\rangle _{k}\right\} $
and $\left\{ \left|0_{L}\right\rangle _{k+1},\left|1_{L}\right\rangle _{k+1}\right\} $
at the end of the entanglement procedure in the same spirit as in
\cite{BPSM08}. To this end, one first prepares auxiliary subnodes
$\left(A_{k}\right)$ and $\left(A_{k+1}\right)$ in the state $\left|\bar{0}_{A}\right\rangle $,
before applying to each pair of subnodes $\left(A_{k},R_{k}\right)$
and $\left(L_{k+1},A_{k+1}\right)$ the sequence of pulses given in
Table \ref{Diagnosis}. At the end of this sequence, the states of
the subnodes $\left(R_{k}\right)$ and $\left(L_{k+1}\right)$ are
unchanged, while the auxiliary subnode $\left(A_{k}\right)$, respectively
$\left(A_{k+1}\right)$, is either in state $\left|\bar{1}_{A}\right\rangle $
if the subnode $\left(R_{k}\right)$ -- respectively $\left(L_{k+1}\right)$,
is singly occupied, or in the state $\left|N_{r_{A}}=1\right\rangle $
if the subnode $\left(R_{k}\right)$ -- respectively $\left(L_{k+1}\right)$,
contains no excitation. One therefore merely needs to selectively
ionize $\left|r_{A}\right\rangle $ in both ensembles $\left(k\right)$
and $\left(k+1\right)$. \emph{Case (A)} If an ion is observed, subnodes
$\left(R_{k}\right)$ and $\left(L_{k+1}\right)$ were not correctly
entangled, they must therefore be reset through state selective ionization,
and the whole procedure in Table \ref{EntanglementGeneration} must
be repeated. \emph{Case (B)} If no ion is observed then one selectively
ionizes the state $\left|1_{A}\right\rangle $ in both ensembles $\left(k\right)$
and $\left(k+1\right)$: \emph{Case (B1)} if an ion is observed, as
expected, the entanglement generation procedure was indeed correctly
performed and the scheme can continue ; \emph{Case (B2)} if no ion
is observed, then most probably an ion detection failed and, to be
sure, the whole procedure for entangling subnodes $\left(R_{k}\right)$
and $\left(L_{k+1}\right)$ must be repeated again, as in the case
(A).

In quite the same way, if a Rydberg atom decays during the entanglement
swapping procedure, an atomic excitation will miss in one of the subnodes
$\left(R_{k},L_{k}\right)_{k=2,\ldots,N-1}$ and the subsequent series
of measurements will therefore fail. In that case, entanglement generation
and swapping should be repeated again after resetting the subnodes
through state selective ionizations.

As seen above, errors can be detected and their effects avoided through
diagnosis and repetition of the erroneous steps. The cost is, however,
an increase of the average time required to run the whole protocol,
which will now be estimated.

Let us first focus on the entanglement generation procedure. As said
in Sec. \ref{SecII}, the success probability for a photon transfer
along a fiber of length $L_{0}$ is $\eta_{t}=\exp\left(-L_{0}/L_{\textrm{att}}\right)$
where $L_{\textrm{att}}\sim22$ km. For $L_{0}=100$ km, $\eta_{t}\simeq1.1\%$.
On the other hand the probability for one Rydberg excited atom to
spontaneously decay during the entanglement generation procedure between
two neighboring subnodes $\left(R_{k},L_{k+1}\right)$ (including
the error diagnosis) is roughly given by $n_{r}\times\frac{\pi}{\Omega}\times\Gamma$
where $n_{r}=23$ is the number of (second-order) $\pi$ pulses involving
Rydberg states, $\Omega$ is the typical value for their Rabi frequency
-- note that these pulses can be either {}``classical-classical''
\emph{i.e.} driven by two laser beams or {}``quantum-classical''
\emph{i.e.} they involve a cavity photon in which case the expression
of the associated Rabi frequency comprises the coupling strength $g_{\pm}$,
and $\Gamma$ is the emission rate of the Ryberg level. Taking for
the parameters the typical values $\Omega=2\pi\times1$MHz and $\Gamma=1$kHz,
one obtains $1-n_{r}\pi\Gamma/\Omega\approx99\%$. Moreover, four
ion detections must be successfully performed (giving either a positive
or negative result) during the diagnosis on the two subnodes $\left(A_{k}\right)$
and $\left(A_{k+1}\right)$. The probability of this event is given
by $\eta_{\textrm{ion}}^{4}\approx96\%$ for the ion detection efficiency
$\eta_{\textrm{ion}}\approx99\%$. Finally, the probability for successfully
entangling a pair of two neighboring subnodes $\left(R_{k},L_{k+1}\right)$
is therefore given by $P_{0}=\eta_{t}\times\left(1-n_{r}\pi\Gamma/\Omega\right)\times\eta_{\textrm{ion}}^{4}\approx1\%$.
Note that the photon transfer is mainly responsible for this low probability,
\emph{i.e.} $P_{0}\approx\exp\left(-L_{0}/L_{\textrm{att}}\right)$;
it is also the longest step of the entanglement generation procedure
since it takes $L_{0}/c=0.5$ms for $L_{0}=100$km and $c=2\times10^{8}$m.s$^{-1}$,
while each pulse takes no more than $\sim1\mu$s, typically, and the
duration of an elementary entanglement generation step is roughly
given by $L_{0}/c$. In average, a pair of neighboring subnodes $\left(R_{k},L_{k+1}\right)$
will be correctly entangled after $1/P_{0}$ repetitions of the entanglement
generation procedure. For the $N$-node chain to be correctly entangled,
the entanglement generation procedure must be repeated on average
a certain number of times $\bar{n}\left(P_{0},N\right)$, whose expression
is calculated in Appendix. For $P_{0}\sim1\%$ and $N=10$ -- \emph{i.e.}
for a total length $L\simeq1000$km, one obtains $\bar{n}\left(P_{0},N\right)\simeq455$.

The same analysis can be achieved for the entanglement swapping step.
The success probability of this step is readily found to be $P_{1}\left(N\right)=\left[\left(1-4\pi\Gamma/\Omega\right)\eta_{\textrm{ion}}^{4}\right]^{N-2}$.
In average, to correctly entangle two remote nodes, $1/P_{1}\left(N\right)$
repetitions of the whole protocol will therefore be necessary. For
$N=10$, $\eta_{\textrm{ion}}\simeq99\%$, $\Omega=2\pi\times1$MHz
and $\Gamma=1$kHz, one gets $P_{1}\left(N\right)\simeq71.3\%$.

Finally, since the time necessary for photon transfer, $L_{0}/c$,
dominates by several orders of magnitude all the other steps, the
average time taken by our protocol can be estimated by $T\sim\frac{L_{0}}{c}\times\frac{\bar{n}\left(P_{0},N\right)}{P_{1}\left(N\right)}$
that is $T\sim0.32$s for the previous set of parameters, to be compared
to the average time it would take via direct transmission through
the lossy optical fiber $\left(1/\chi_{r}\right)\times\left(1/\exp^{-L/L_{att}}\right)\sim5.5\times10^{9}$s,
where $\chi_{r}$ is the repetition rate of the source of photons
which we took equal to $10$ Ghz for our estimation.

To conclude this section, let us point out that other errors can affect
our protocol. First, Rydberg levels can be multiply excited due to
the finite value of $\Delta_{dd}$. They constitute losses for our
protocol, just as spontaneous emission, and are therefore already
dealt with by the scheme. Their probability is $\left(\frac{\Omega}{\Delta_{dd}}\right)^{2}\lesssim1\%$
and only very weakly modifies $P_{0}$ and $P_{1}$.

Secondly, uncertainty in the number of atoms in the sample may lead
to inaccuracies in the Rabi frequencies. It was, however, noted in
\cite{ZMHZ10}, that such errors can be made as low as $1\%$ ; moreover,
as suggested in \cite{BPSM08}, they can also be dealt with by composite
pulse techniques. We shall not consider them here.

\begin{table}
\begin{centering}
\begin{tabular}{|l|}
\hline 
\emph{i}) $\pi$ pulse : $\left|1_{L,R}\right\rangle \leftrightarrow\left|r_{+}\right\rangle $\tabularnewline
\hline 
\emph{ii})$\pi$ pulse : $\left|0_{A}\right\rangle \leftrightarrow\left|r_{A}\right\rangle $\tabularnewline
\hline 
\emph{iii}) $\pi$ pulse : $\left|r_{A}\right\rangle \leftrightarrow\left|1_{A}\right\rangle $\tabularnewline
\hline 
\emph{iv}) $\pi$ pulse : $\left|r_{+}\right\rangle \leftrightarrow\left|1_{L,R}\right\rangle $\tabularnewline
\hline 
\emph{v}) $\pi$ pulse : $\left|0_{L,R}\right\rangle \leftrightarrow\left|r_{-}\right\rangle $\tabularnewline
\hline 
\emph{vi}) $\pi$ pulse : $\left|0_{A}\right\rangle \leftrightarrow\left|r_{A}\right\rangle $\tabularnewline
\hline 
\emph{vii}) $\pi$ pulse : $\left|r_{A}\right\rangle \leftrightarrow\left|1_{A}\right\rangle $\tabularnewline
\hline 
\emph{viii}) $\pi$ pulse : $\left|r_{-}\right\rangle \leftrightarrow\left|0_{L,R}\right\rangle $\tabularnewline
\hline 
\end{tabular}
\par\end{centering}

\caption{Pulse sequence for diagnosing errors occurring during entanglement
generation procedure.}

\label{Diagnosis} 
\end{table}

\section{A physical implementation}

\begin{figure*}
\begin{centering}
\includegraphics[width=14cm]{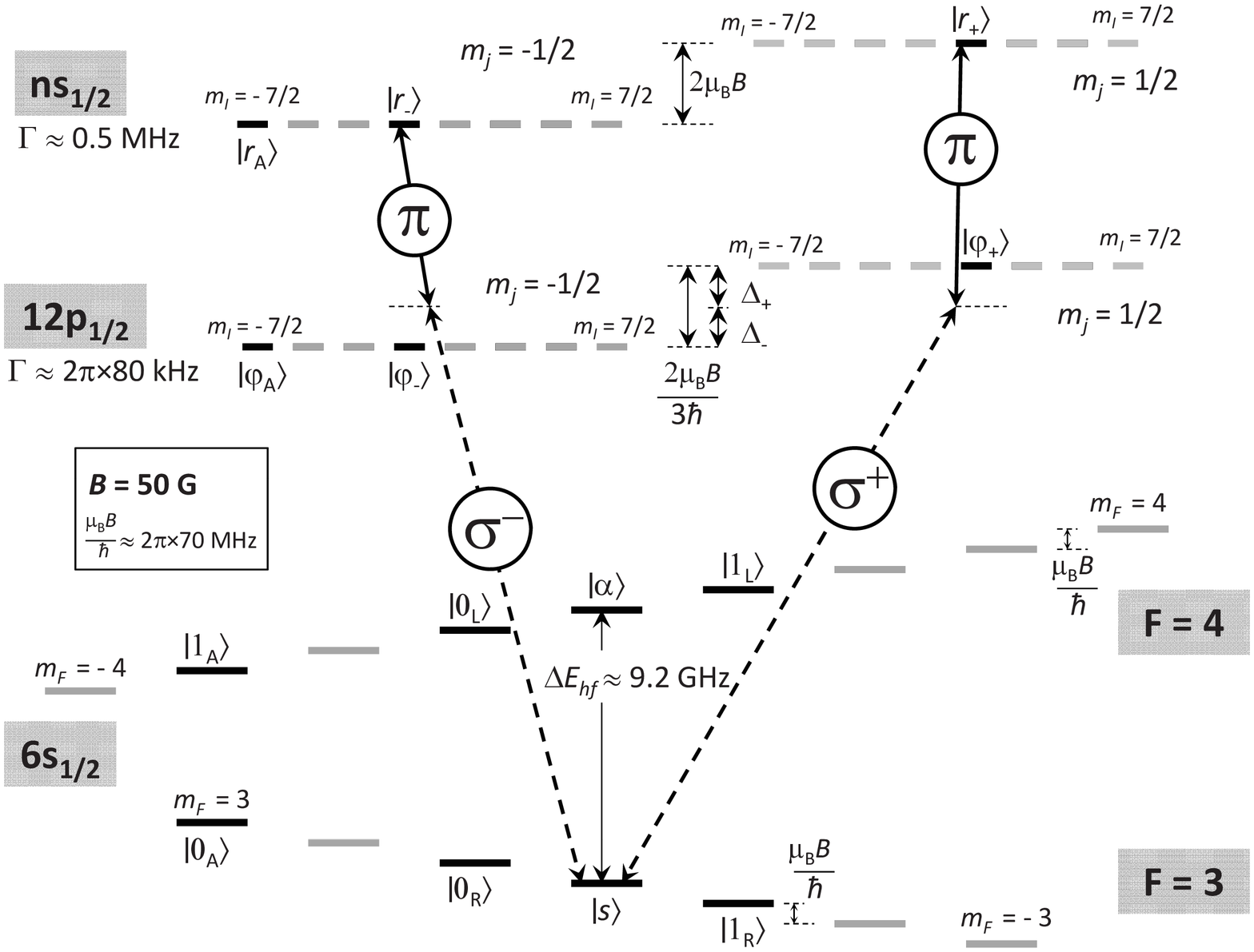} 
\par\end{centering}

\caption{Physical implementation of our scheme with Cs atoms. The figure shows
the states $\left|j_{l}\right\rangle _{j=0,1;l=L,R,A},\left|\varphi_{+,-,A}\right\rangle $
and $\left|r_{+,-,A}\right\rangle $. For sake of clarity, the figure
only shows the quantum-classical paths $\left|r_{\pm}\right\rangle \leftrightarrow\left|\varphi_{\pm}\right\rangle \leftrightarrow\left|s\right\rangle $
used for converting an atomic excitation into a cavity photon and
vice-versa. The dashed-line arrows stand for the transitions driven
by the cavity modes $\pm$, of respective polarizations $\sigma_{\pm}$,
the full lines stand for the transitions driven by laser beams of
linear polarization. Note that unwanted transitions involving relevant
states are out of resonance and therefore highly suppressed. }

\label{Fig2} 
\end{figure*}

In this section, we suggest a physical implementation of our scheme
with ensembles of Cs atoms, placed in linear cavities. Fig. \ref{Fig2}
presents a possible choice for the internal states used in our protocol.
The reservoir state $\left|s\right\rangle $ and the subnode states
$\left|0_{L,A,R}\right\rangle ,\left|1_{L,A,R}\right\rangle $ correspond
to different hyperfine components of the ground level $6s_{1/2},\: I=7/2$
\begin{eqnarray*}
\left|s\right\rangle  & \equiv & \left|F=3;m_{F}=0\right\rangle \\
\left|0_{R}\right\rangle  & \equiv & \left|F=3;m_{F}=1\right\rangle \\
\left|0_{L}\right\rangle  & \equiv & \left|F=4;m_{F}=-1\right\rangle \\
\left|1_{R}\right\rangle  & \equiv & \left|F=3;m_{F}=-1\right\rangle \\
\left|1_{L}\right\rangle  & \equiv & \left|F=4;m_{F}=1\right\rangle \\
\left|0_{A}\right\rangle  & \equiv & \left|F=3;m_{F}=3\right\rangle \\
\left|1_{A}\right\rangle  & \equiv & \left|F=4;m_{F}=-3\right\rangle 
\end{eqnarray*}
They are coupled to the Rydberg states $\left|r_{\pm}\right\rangle =\left|ns_{1/2},m_{j}=\pm1/2,m_{I}=\pm1/2\right\rangle ,\left|r_{A}\right\rangle =\left|ns_{1/2},m_{j}=-1/2,m_{I}=-7/2\right\rangle $,
with $n\sim70$, via the intermediate states $\left|\varphi_{\pm}\right\rangle =\left|12p_{1/2},m_{j}=\pm1/2,m_{I}=\pm1/2\right\rangle $
and $\left|\varphi_{A}\right\rangle =\left|12p_{1/2},m_{j}=-1/2,m_{I}=-7/2\right\rangle $.
Note that for the intermediate and Rydberg levels, the hyperfine structure
may be neglected, which legitimates the use of the decoupled basis.
We assume the availability of light sources and cavities at the wavelengths
of the required transitions -- \emph{i.e.} $335$nm and $6.5\mu$m
for the $6s\leftrightarrow12p$ and $12p\leftrightarrow75s$ transitions,
respectively \footnote{If the $\left|6s_{1/2}\right\rangle  \leftrightarrow \left|12p_{1/2}\right\rangle$ transition is driven via the $\left|7p_{1/2}\right\rangle$ state with a classical field at $460$nm, the quantum field at the upper transition is at $1.24 \mu$m which conveniently matches
telecommunication fibers.}. Selection rules show that almost all the transitions
necessary to our scheme are \emph{allowed} and that, in particular,
the transitions $\left|\varphi_{\pm}\right\rangle \rightarrow\left|s\right\rangle $
have different polarizations, that is $\sigma^{+}$ and $\sigma^{-}$,
as required (Note that, by setting the quantization axis, \emph{i.e.}
the direction of the applied magnetic field, along the axis of the
linear cavity, one highly supresses modes with $\pi$ polarization).
Only the direct coupling\textbf{ }$\left|0_{L,R,A}\right\rangle \leftrightarrow\left|1_{L,R,A}\right\rangle $
is not permitted. To overcome this difficulty, we suggest to resort
to intermediate states. To be more explicit, to apply a unitary transformation
in the subspace $\left\{ \left|0_{A}\right\rangle ,\left|1_{A}\right\rangle \right\} $
we propose to first transfer the population from $\left|0_{A}\right\rangle $
to \textbf{$\left|\varphi_{A}\right\rangle $}, then to run the desired
transformation between \textbf{$\left|\varphi_{A}\right\rangle $}
and \textbf{$\left|1_{A}\right\rangle $} -- which are indeed coupled,
and finally transfer the population back from\textbf{ }$\left|\varphi_{A}\right\rangle $
to $\left|0_{A}\right\rangle $. The same trick can be used to emulate
a coupling between $\left|0_{R,L}\right\rangle $ and $\left|1_{R,L}\right\rangle $
: one first transfers the population from $\left|0_{R,L}\right\rangle $
to $\left|\varphi_{-}\right\rangle $ then to $\left|\alpha\right\rangle =\left|6s_{\frac{1}{2}};I=\frac{7}{2};F=4;m_{F}=0\right\rangle $
then to $\left|\varphi_{+}\right\rangle $ ; one then applies the
desired transformation between $\left|\varphi_{+}\right\rangle $
and $\left|1_{L,R}\right\rangle $ before transferring the population
back from $\left|\varphi_{+}\right\rangle $ to $\left|0_{R,L}\right\rangle $
along the same path as in the first step.\textbf{ }

Moreover, as indicated on Fig. \ref{Fig2}, a magnetic field is applied
to lift the degeneracies of the different levels. The specific choice
we made, that is $B\sim50$G, results in splittings of $\frac{\mu_{B}B}{4\hbar}\sim2\pi\times17.5\mbox{MHz}$,
$\frac{2\mu_{B}B}{3\hbar}\sim2\pi\times47\mbox{MHz}$ for the hyperfine
components $\left|F=3,4;m_{F}\right\rangle $ of the ground level
and the excited states \textbf{$\left|\varphi_{\pm}\right\rangle ,\left|r_{\pm}\right\rangle $},
respectively. These splittings assure that the different transitions
required by our protocol are \emph{selectively addressable}, provided
that the relevant effective two-photon coupling constant is smaller
than $\frac{\mu_{B}B}{\hbar}\sim2\pi\times70$MHz and respects the
finite lifetime of the Rydberg level; moreover, the detunings from
the intermediate states $\left|\varphi_{\pm}\right\rangle $ must
be chosen larger than $\gamma_{12p}=\tau^{-1}\sim2\pi\times80$kHz,
$\tau$ being the lifetime of the level $12p_{1/2}$. Typical values
of $\Omega\sim2\pi\times1\mbox{MHz}$ and $\Delta\sim2\pi\times10\mbox{MHz}$
fulfill the previous requirements, and can be achieved for both {}``classical-classical''
and {}``quantum-classical'' paths in a sample of a few hundreds
of atoms with $\Omega_{laser}\sim2\pi\times1$ MHz and $g_{\pm}\sim2\pi\times0.1$
MHz. Note that the size of the samples should also be small enough
so as to remain in the full blockade regime. As shown in \cite{BMS07},
a cloud of $\sim5\mu\mbox{m}$ of a few hundreds of atoms, exhibit
Rydberg dipole-dipole interactions of at least $\Delta_{dd}\sim2\pi\times100\mbox{MHz}$,
which is indeed much larger than $\Omega$ and therefore efficiently
forbids multiple Rydberg excitations. Finally, all the two-photon
processes required in our protocol, including gates on the quantum
register, can be performed on the $\mu\mbox{s}$ timescale. Finally,
note that the spontaneous emission from the $12p_{1/2}$ does not
constitute a problem when performing unitaries in the subspace $\left\{ \left|0_{A}\right\rangle ,\left|1_{A}\right\rangle \right\} $
through actually populating the state \textbf{$\left|\varphi_{A}\right\rangle $}.
Indeed, in this case, we must only fulfill the condition $\Omega_{laser}\ll\frac{\mu_{B}B}{\hbar}\sim2\pi\times70$MHz
to ensure that no unwanted transition such as $\left|0_{A}\right\rangle \leftrightarrow\left|12p_{1/2},m_{j}=-1/2,m_{I}=-3/2\right\rangle $
take place. The frequencies of all the required manipulations -- transfers
$\left|0_{A}\right\rangle \leftrightarrow\left|\varphi_{A}\right\rangle $
and unitaries $\left|1_{A}\right\rangle \leftrightarrow\left|\varphi_{A}\right\rangle $,
can therefore be taken as large as $\sim2\pi\times7$ MHz, while the
decay rate of the level $12p_{1/2}$ is only $\sim2\pi\times80$kHz
: the whole unitary process can therefore be run before the decay
of the level $12p_{1/2}$ plays any role.

\section{Discussion}

We now summarize the main differences between our scheme and the most
recent works on the subject \cite{ZMHZ10,HHHLS10}. As noted in \cite{ZMHZ10,HHHLS10},
the use of Rydberg blocked ensembles allows one to perform entanglement
swapping via deterministic manipulations and not through probabilistic
photonic detections. The main originality of our proposal is that
we also completely got rid of photonic detections during the linking
procedure. Indeed, here, photons are simply transmitted from one site
and reabsorbed by its neighbor in an efficient and faithful way. The
heralded linking is performed only via deterministic ensemble manipulations
and ion detections, whose efficiency can be made very close to one.
In that respect, our scheme is more of a relay type, as defined in
\cite{SSRG09}. It is also important to note that, contrary to \cite{ZMHZ10},
we do not need different ensembles for encoding what we called {}``subnodes''
in the present work, but use the multilevel structure of the atomic
spectrum to store three subnodes in the same ensemble. It therefore
means that we do not rely on Rydberg blockade between two ensembles,
a rather challenging task. We also note that another proposal for
a quantum repeater based on atomic ensembles was put forward in \cite{BJG10}.
There, however, the authors did not rely on Rydberg blockade phenomenon
but rather on fluoresecence detection of excitations stored in the
atoms after low intensity laser excitation and Raman scattering.

\begin{figure}
\begin{centering}
\includegraphics[width=8cm]{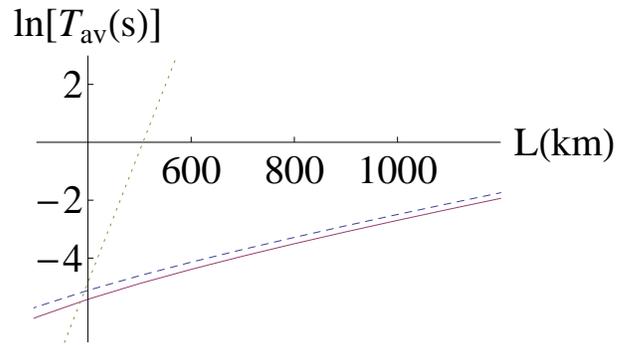} 
\par\end{centering}

\caption{(Color online) Logarithm of the average time required for entangling
two remote nodes over the distance $L$ (km), for a total number of
nodes $N=2^{4}=16$, via the direct exchange of a photon (dotted line),
through the repeaters described in \cite{SWM10,HHHLS10} (dashed line),
and through our protocol (full line). }

\label{Fig3} 
\end{figure}

Finally it is worth comparing the total average time needed by our
protocol to that required by other schemes, as a figure of merit.
Fig. \ref{Fig3} displays the logarithm of the average time necessary
for entangling two remote nodes by the direct exchange of a photon
-- a generous repetition rate of $10$ GHz for the photon source was
assumed, by the protocol described in \cite{ZMHZ10,HHHLS10}, and
by our protocol as functions of the distance $L$ between the two
nodes to entangle, for a fixed number of $N=2^{4}=16$ nodes. As in
\cite{ZMHZ10,HHHLS10}, a photodetection efficiency of $\eta_{\mbox{pd}}\simeq0.9$
and a retrieval efficiency of $\eta_{r}\simeq0.9$ (in our cavity
model, this retrieval efficiency was taken equal to one) were assumed.
It appears that our protocol is quicker, though asymptotically equivalent
for $L\rightarrow\infty$, which is explained by our assumption of
the ion detection efficiency exceeding that of photons.

\section{Conclusion}

In this paper we proposed a quantum repeater scenario based on Rydberg
blocked ensembles placed in cavities which are linked by optical fibers.
Entanglement generation between two neighboring nodes is performed
in a heralded way by the transmission of a photon whose polarization
is entangled with the state of the first atomic ensemble, followed
by its absorption by the neighboring atomic ensemble. Photon losses
and spontaneous emission from the Rydberg level can be detected thanks
to an error-syndrome measurement involving ensemble laser manipulations,
ionizing pulses and (very efficient) ion detections. An implementation
with Cs atoms was suggested and analyzed.

Contrary to protocols previously proposed, the scheme presented here
does not make use of any (inefficient) photodetection: this potentially
allows for a speedup in the entanglement generation, as confirmed
by numerical simulations. Finally main error sources were analyzed.
Future work should be devoted to a closer investigation of the practical
feasibility of our scheme with real cavities and fibers. 
\begin{acknowledgments}
E. B. thanks M. Raoult, Jean-Louis Le Gouet and F. Prats for fruitful discussions.
\end{acknowledgments}
\appendix

\section{Derivation of the expression of the average number of steps in entanglement
generation\textup{\normalsize \label{Appendix}}}

As described in Sec. \ref{sec:The-scheme}, the entanglement generation
procedure is performed in two steps. During each of these steps, $K=N/2$
subnodes are entangled by pairs. A pair of subnodes is correctly entangled
with the probability $P_{0}$.

Let us first compute the probability $p_{K}\left(n\right)$ that $K$
pairs are correctly entangled within exactly $n$ repetitions. This
means that, within the $\left(n-1\right)$ first steps, at least one
pair is not entangled. Considering all the possible cases, one establishes
the following recurrence formula 
\begin{eqnarray*}
p_{K}\left(n\right) & = & p_{K-1}\left(n\right)p_{1}\left(n\right)+p_{K-1}\left(n\right)\left(\sum_{m=1}^{n-1}p_{1}\left(m\right)\right)\\
 &  & +\left(\sum_{m=1}^{n-1}p_{K-1}\left(m\right)\right)p_{1}\left(n\right)\\
 & = & p_{K-1}\left(n\right)\left(\sum_{m=1}^{n}p_{1}\left(m\right)\right)+\left(\sum_{m=1}^{n-1}p_{K-1}\left(m\right)\right)p_{1}\left(n\right)
\end{eqnarray*}
 and, noting that $p_{1}\left(n\right)=P_{0}\left(1-P_{0}\right)^{n-1}$
one gets 
\begin{eqnarray}
p_{K}\left(n\right) & = & p_{K-1}\left(n\right)\left[1-\left(1-P_{0}\right)^{n}\right]\label{rec}\\
 &  & +\left(\sum_{m=1}^{n-1}p_{K-1}\left(m\right)\right)P_{0}\left(1-P_{0}\right)^{n-1}.
\end{eqnarray}
 Setting $S_{K}\left(n\right)\equiv\sum_{m=1}^{n}p_{K}\left(m\right)$,
one derives from Eq. (\ref{rec}) the relation $S_{K}\left(n\right)=S_{K-1}\left(n\right)\left[1-\left(1-P_{0}\right)^{n}\right]$
whence $S_{K}\left(n\right)=S_{1}\left(n\right)\left[1-\left(1-P_{0}\right)^{n}\right]^{K-1}$
and, since $S_{1}\left(n\right)=\left[1-\left(1-P_{0}\right)^{n}\right]$,
$S_{K}\left(n\right)=\left[1-\left(1-P_{0}\right)^{n}\right]^{K}$.
One finally deduces the expression for $p_{K}\left(n\right)$ from
$p_{K}\left(n\right)=S_{K}\left(n\right)-S_{K}\left(n-1\right)$ 
\[
p_{K}\left(n\right)=\left[1-\left(1-P_{0}\right)^{n}\right]^{K}-\left[1-\left(1-P_{0}\right)^{n-1}\right]^{K}.
\]
 The average number of repetitions one needs to entangle the K pairs
is simply given by $\sum_{n=1}^{+\infty}np_{K}\left(n\right)$.

\begin{figure}
\begin{centering}
\includegraphics[width=9cm]{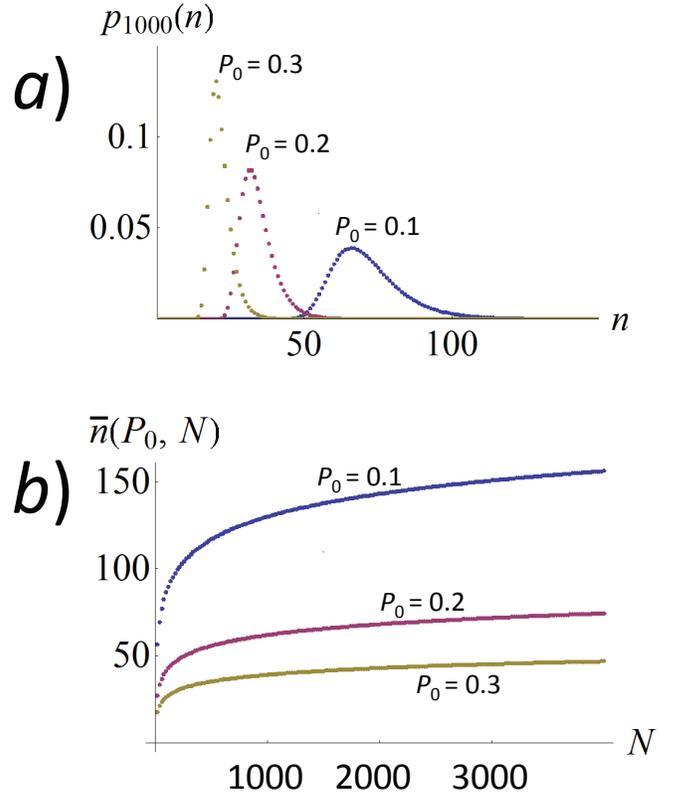} 
\par\end{centering}

\caption{(Color online) a) Generic behaviour of the probability $p_{K}\left(n\right)$
for $K=1000$ and three different values of $P_{0}=0.1,0.2,0.3$.
The probability is a peaked curve around its maximum $n_{K}^{max}$.
b) Behaviour of $\bar{n}\left(P_{0},N\right)$ as a function of $N$
for three different values of $P_{0}=0.1,0.2,0.3$. }

\label{FigAp1} 
\end{figure}

Since the entanglement of the two groups of $K=N/2$ pairs of subnodes
$\left(R_{2k},L_{2k+1}\right)$ and $\left(R_{2k+1},L_{2\left(k+1\right)}\right)$
is performed independently and successively, the total average number
of repetitions required is simply $\bar{n}\left(P_{0},N\right)=2\sum_{n=1}^{+\infty}np_{K}\left(n\right)$
represented on Fig. \ref{FigAp1}, or, more explicitly 
\begin{eqnarray}
\bar{n}\left(P_{0},N\right) & = & 2\sum_{n=1}^{+\infty}n\left\{ \left[1-\left(1-P_{0}\right)^{n}\right]^{K}\right.\nonumber \\
 &  & \left.-\left[1-\left(1-P_{0}\right)^{n-1}\right]^{K}\right\} .\label{nav}
\end{eqnarray}

\begin{figure}
\begin{centering}
\includegraphics[width=9cm]{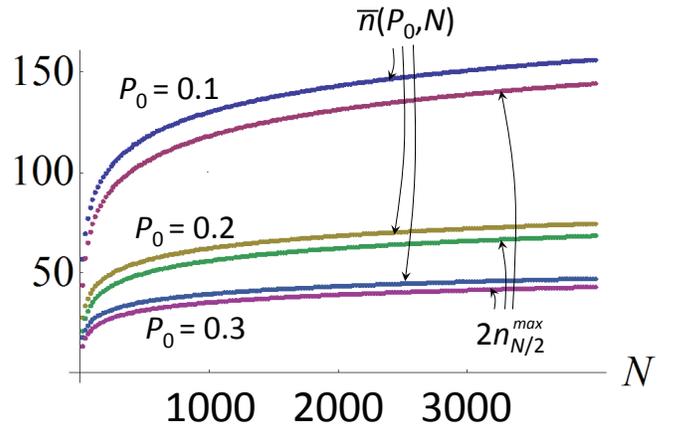} 
\par\end{centering}

\caption{(Color online) Comparison of $\bar{n}\left(P_{0},N\right)$and $2n_{N/2}^{max}$
for three different values of $P_{0}=0.1,0.2,0.3$. }

\label{FigAp2} 
\end{figure}

Let us now derive a simple lower bound for $\bar{n}\left(P_{0},N\right)$.
We shall first note that $p_{K}\left(n\right)=f_{K}\left(n\right)-f_{K}\left(n-1\right)$
where $f_{K}\left(n\right)=\left[1-(1-P_{0})^{n}\right]^{K}$ ; for
$n\gg1$, one thus has $p_{K}\left(n\right)\simeq\frac{df_{K}}{dn}\left(n\right)$
and therefore one can calculate the approximate position of the maximum
of $p_{K}\left(n\right)$ by deriving $\frac{df_{K}}{dn}\left(n\right)$.
Doing so, one obtains a maximum for $p_{K}\left(n\right)$ at $n=n_{K}^{max}=-\ln K/\ln\left(1-P_{0}\right)$.
As can be seen on Fig. \ref{FigAp1}, the distribution $p_{K}\left(n\right)$
is not symmetric around its maximum : the position of its peak therefore
cannot, strictly speaking, be identified with $\bar{n}\left(P_{0},N=2K\right)/2$.
It, however, gives a good order of magnitude, $\bar{n}\left(P_{0},N\right)\gtrsim-2\ln\left(N/2\right)/\ln\left(1-P_{0}\right)$,
as can be checked on Fig. \ref{FigAp2}. In particular, the expression
of $2n_{N/2}^{max}$ gives a good indication on how $\bar{n}\left(L_{0},N\right)$
scales with the physical parameters.

\end{document}